\documentclass[twocolumn,showpacs,floatfix,superscriptaddress,amsmath,amssymb,prl]{revtex4}
\usepackage{amsmath}
\usepackage{amsfonts}
\usepackage{bbm}
\usepackage{mathrsfs}
\usepackage{txfonts}
\usepackage{amssymb}
\usepackage{graphicx}
\usepackage{hyperref}
\usepackage{ulem}
 \usepackage{overpic}
 \usepackage{psfrag}
\usepackage{tabularx}
\usepackage{array}
\usepackage{placeins}
\makeatletter
\renewcommand\subsection{\@startsection
{subsection}{2}{0mm}
 {-\baselineskip}
 {0.5\baselineskip}
{\FloatBarrier\normalfont\Large\bfseries}}
\makeatother
\newcommand{\be}{\begin{equation}}
\newcommand{\ee}{\end{equation}}

\newcommand{\PreserveBackslash}[1]{\let\temp=\\#1\let\\=\temp}

\newcommand{\ket} [1] {| #1 \rangle}

\begin{document}
\title{Tensor network states for quantum spin ladders}

\author{Sheng-Hao Li} \affiliation{Centre for Modern Physics and Department of Physics,
Chongqing University, Chongqing 400044, The People's Republic of
China}

\author{Yao-Heng Su} \affiliation{Centre for Modern Physics and Department of Physics,
Chongqing University, Chongqing 400044, The People's Republic of
China}

\author{Yan-Wei Dai} \affiliation{Centre for Modern Physics and Department of Physics,
Chongqing University, Chongqing 400044, The People's Republic of
China}

\author{Huan-Qiang Zhou} \affiliation{Centre for Modern Physics and Department of
Physics, Chongqing University, Chongqing 400044, The People's
Republic of China}

\begin{abstract}
We have developed an efficient tensor network algorithm for spin
ladders, which generates ground-state wave functions for
infinite-size quantum spin ladders. The algorithm is able to
efficiently compute the ground-state fidelity per lattice site, a
universal phase transition marker, thus offering a powerful tool to
unveil quantum many-body physics underlying spin ladders.  To
illustrate our scheme, we consider the two-leg and three-leg
Heisenberg spin ladders with staggering dimerization. The
ground-state phase diagram thus yielded is reliable, compared with
the previous studies based on the density matrix renormalization
group. Our results indicate that the ground-state fidelity per
lattice site successfully captures quantum criticalities in spin
ladders.
\end{abstract}
 \maketitle
{\it Introduction.} Tensor networks (TN)  provide a convenient means
to represent quantum wave functions in classical simulations of
quantum many-body lattice systems, such as the matrix product states
(MPS)~\cite{b6,b7,b8,TEBD,iTEBD} in one spatial dimension and the
projected entangled-pair state (PEPS)~\cite{b10,PEPS,iPEPS} in two
and higher spatial dimensions. The development of various numerical
algorithms in the context of the TN representations has led to
significant advances in our understanding of quantum many-body
lattice systems in both one and two spatial
dimensions~\cite{b8,TEBD,iTEBD,b10,PEPS,iPEPS,b11,onetwo1,onetwo2,onetwo3,onetwo4,onetwo5,onetwo6,onetwo7,onetwo8}.
Lying between quantum lattice systems in one and two spatial
dimensions, spin ladders have attracted a lot of attention, due to
their intriguing critical properties. Given the importance of spin
ladder systems in condensed matter physics, it is somewhat
surprising that no efforts have been made to develop any efficient
algorithm in the context of the TN representations.

This paper aims to fill in this gap.  The algorithm generates
efficiently ground-state wave functions for infinite-size spin
ladders. In addition, it allows to efficiently compute the
ground-state fidelity per lattice site, a universal phase transition
marker, thus offering a powerful tool to unveil quantum many-body
physics underlying spin ladders. In fact, as argued in
Refs.~\cite{b1,b2,b3,b4,b5,whl}, the ground-state fidelity per
lattice site is able to capture drastic changes of the ground-state
wave functions around a critical point. To illustrate our scheme, we
consider the two-leg and three-leg Heisenberg spin ladders with
staggering dimerization. The ground-state phase diagram thus yielded
is reliable, compared with the previous
studies~\cite{dimer,QCriticality3} based on the density matrix
renormalization group (DMRG)~\cite{DMRG}. Our results indicate that
the ground-state fidelity per lattice site successfully captures
quantum criticalities in spin ladders.

{\it Tensor network representation for spin ladders.} Let us
describe the TN representation suitable to describe a ground-state
wave function for an infinite-size spin ladder. Suppose the
Hamiltonian is translationally invariant under shifts by either one
or two lattice sites along the legs: $H=\sum_{\langle i, \alpha
\rangle}h_{\langle i, \alpha \rangle}$, with  the $\langle i, \alpha
\rangle$-th plaquette Hamiltonian density $h_{\langle i, \alpha
\rangle}$ acting on sites $i$ and $(i+1)$ along the $\alpha$-th and
$(\alpha+1)$-th legs. Here, $\langle i, \alpha \rangle$ runs over
all the possible plaquettes by taking $i \in \{-\infty,
\cdots,+\infty \}$, and $\alpha =1, \cdots, n-1$, with $n$ being the
number of the legs. Assume that the TN representation for a wave
function enjoys the translational invariance under shifts by two
lattice sites along the legs. In the following, we focus on a
detailed description for a two-leg spin ladder, with a brief
discussion for an $n$-leg ladder system.

\begin{figure}
\begin{center}
\includegraphics[angle=90,width=2.8in]{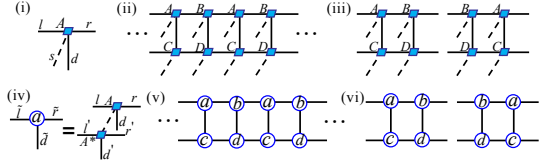}
\end{center}
\vspace*{-0.5cm}
 \caption{(Color online) (i) Four-index tensor
$A^{s}_{lrd}$ used to represent a TN representation for the
ground-state wave function for an infinite-size two-leg spin ladder,
with $s$ being a physical index, $l$, $r$, and $d$ denoting the
inner indices. (ii) The pictorial representation for a TN state
$\ket{\psi}$ with leg and rung bonds, which are used to absorb an
operator acting on the $i$-th plaquette. (iii) Two different choices
of the unit cell for an infinite TN state, made of four four-index
tensors $A$, $B$, $C$, and $D$. (iv) A double tensor
$a_{\tilde{\ell} \tilde{r}\tilde{d}}$ is formed from the four-index
tensor $A_{\ell rd}^{s}$ and its complex conjugate ${(A^*)_{\ell^{'}
r^{'}d^{'}}^{s}}$, with $\tilde{\ell}\equiv(\ell,\ell^{'})$,
$\tilde{r}\equiv(r,r^{'})$, and $\tilde{d}\equiv(d,d^{'})$. (v) The
TN representation for the norm of a ground-state wave function in an
infinite-size spin ladder. (vi) Two different choices of the unit
cells for the norm tensor net work, consisting of four double
tensors $a$, $b$, $c$, and $d$.}\label{FIG1}
\end{figure}

For an infinite-size two-leg spin ladder system,  we need only four
different four-index tensors $A_{\ell rd}^{s}$, $B_{\ell rd}^{s}$,
$C_{\ell ru}^{s}$, and $D_{\ell ru}^{s}$ to store the wave function.
Here, $A_{\ell rd}^{s}$, $B_{\ell rd}^{s}$, $C_{\ell ru}^{s}$, and
$D_{\ell ru}^{s}$ are made of complex numbers labeled by one
physical index $s$ and four inner bond indices $\ell$, $r$, $u$ and
$d$, where $s=1,...,\mathbbm{d}$, with $\mathbbm{d}$ being the
dimension of the local Hilbert space, and $\ell$, $r$, $u$,
$d=1,...,\mathbb{D}$, with $\mathbb{D}$ being the bond dimension.  A
four-index tensor $A_{\ell rd}^{s}$ is visualized in
Fig.\ref{FIG1}(i), with a similar pictorial representation for the
tensors $B_{\ell rd}^{s}$, $C_{\ell ru}^{s}$, and $D_{\ell ru}^{s}$.
A TN representation for the ground-state wave function is shown for
an infinite-size two-leg spin ladder in Fig.\ref{FIG1}(ii). There
are two different but equivalent choices of the unit cell for such
an infinite-size TN: one is chosen as $A$, $B$, $D$, and $C$
clockwise if $i$ is even, the other is chosen as $B$, $A$, $C$, and
$D$ clockwise if $i$ is odd, see Fig.\ref{FIG1}(iii).

Now let us turn to the computation of the norm for a quantum state
wave function. To this end, we introduce double tensors
$a_{\tilde{\ell} \tilde{r}\tilde{d}}$, $b_{\tilde{\ell}
\tilde{r}\tilde{d}}$, $c_{\tilde{\ell} \tilde{r}\tilde{u}}$, and
$d_{\tilde{\ell} \tilde{r}\tilde{u}}$, with
$\tilde{\ell}\equiv(\ell,\ell^{'})$, $\tilde{r}\equiv(r,r^{'})$,
$\tilde{u}\equiv(u,u^{'})$, and $\tilde{d}\equiv(d,d^{'})$. They
form from the four-index tensors $A_{\ell rd}^{s}$, $B_{\ell
rd}^{s}$, $C_{\ell ru}^{s}$, and $D_{\ell ru}^{s}$,  and their
complex conjugates, see Fig.\ref{FIG1}(iv) for a pictorial
representation of the double tensors. With these double tensors, the
TN for the norm of a wave function is shown in Fig.\ref{FIG1}(v).
Again, we have two different but equivalent choices for the unit
cell of the norm TN: one is $a$, $b$, $d$, and $c$ clockwise if $i$
is even, the other is $b$, $a$, $c$, and $d$ clockwise if $i$ is
odd, see Fig.\ref{FIG1}(vi).

\begin{figure}
\begin{center}
\includegraphics[angle=90,width=2.8in]{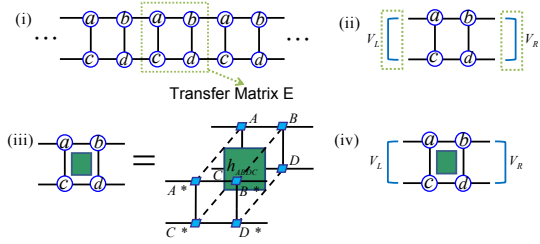}
\end{center}
\vspace*{0cm}
 \caption{(Color online) The ground-state energy per unit cell is computed
 for translation-invariant spin ladders. (i) The transfer matrix $E$
 for an infinite-size norm tensor network, which is constructed
from four double tensors $a_{\tilde{\ell} \tilde{r}\tilde{d}}$,
$b_{\tilde{\ell} \tilde{r}\tilde{d}}$, $c_{\tilde{\ell}
\tilde{r}\tilde{u}}$, and $d_{\tilde{\ell} \tilde{r}\tilde{u}}$,
with $\tilde{\ell}\equiv(\ell,\ell^{'})$,
$\tilde{r}\equiv(r,r^{'})$, $\tilde{u}\equiv(u,u^{'})$, and
$\tilde{d}\equiv(d,d^{'})$. (ii) The dominant left and right
eigenvectors $V_{L}$ and $V_{R}$ of the transfer matrix $E$. (iii) A
unit cell with the Hamiltonian density $h_{ABDC}$ acted on the
plaquette. (iv) The ground-state energy per unit cell is computed
from the eigenvectors $V_L$, $V_R$, four four-index tensors $A_{\ell
rd}^{s}$, $B_{\ell rd}^{s}$, $C_{\ell ru}^{s}$, and $D_{\ell
ru}^{s}$.}\label{FIG2}
\end{figure}

The expectation value of an operator acting on a plaquette, such as
the ground-state energy per unit cell, also admits a TN
representation, which absorbs the operator acting on a plaquette for
an infinite-size spin ladder system. For a randomly chosen initial
state $\ket{\psi_0}$, the energy is expressed as,
\begin{equation}
E=\frac{\langle \psi_0 | H |\psi_0 \rangle}{\langle \psi_0 |\psi_0
\rangle}.
\end{equation}
For different choices of the unit cell, we get two different but
equivalent forms of the zero-dimensional transfer matrix $E$
constructed from four double tensors $a_{\tilde{\ell}
\tilde{r}\tilde{d}}$, $b_{\tilde{\ell} \tilde{r}\tilde{d}}$,
$d_{\tilde{\ell} \tilde{r}\tilde{u}}$, and $c_{\tilde{\ell}
\tilde{r}\tilde{u}}$, one of them is shown in Fig.~\ref{FIG2}(i).
The dominant left and right eigenvectors of the transfer matrix $E$
constitute the environment tensors, visualized in
Fig.~\ref{FIG2}(ii). This enables us to absorb an operator acting on
the $i$-th plaquette $A$, $B$, $D$, and $C$ clockwise, if  $i$ is
even, as shown in Fig.~\ref{FIG2}(iii), and compute the energy per
unit cell, as shown in Fig.~\ref{FIG2}(iv). The same procedure may
be used to compute the energy per  unit cell for an operator acting
on the $i$-th plaquette $B$, $A$, $C$, and $D$ clockwise, if $i$ is
odd.

\begin{figure}
\begin{center}
\includegraphics[angle=90,width=3.0in]{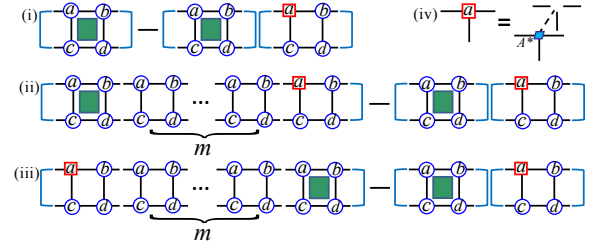}
\end{center}
\vspace*{0cm} \caption{(Color online) The contribution to the energy
gradient  for an infinite-size spin ladder consists of three parts:
(i) the hole cell with the tensor $A_{\ell rd}^{s}$ removed and the
Hamiltonian cell with the Hamiltonian density acting on a plaquette
locate on the same cell; (ii) the hole cell locates on the right
hand side of the Hamiltonian cell; (iii) the hole cell locates on
the left hand side of the Hamiltonian cell. In the latter two cases,
there are $m$ cells between the hole and Hamiltonian cells, where $m
\in (0,1,2,3,\cdots)$. Here, the hole cell is visualized in (iv),
with the tensor  $A_{\ell rd}^{s}$ removed.}\label{FIG3}
\end{figure}

To update the TN representation, we compute the energy gradient with
respect to four-index tensors:
\begin{align}
\frac{\partial E}{\partial A_{\ell rd}^{s}}=\frac{{\partial \langle
\psi_0 | H |\psi_0 \rangle}/{\partial A_{\ell rd}^{s}}}{\langle
\psi_0 |\psi_0 \rangle}- E \cdot \frac{{\partial \langle \psi_0
|\psi_0 \rangle}/{\partial A_{\ell rd}^{s}}}{\langle \psi_0 |\psi_0
\rangle}.
\end{align}
Here, a four-index tensor $A_{\ell rd}^{s}$ is used to explain how
to efficiently evaluate the energy gradient in the context of the
tensor network representation for an infinite-size two-leg spin
ladder, with the details visualized in Fig.~\ref{FIG3}. Notice that
the contributions to the energy gradient come from three parts: (i)
the hole cell with the four-index tensor $A_{\ell rd}^{s}$ absent
and the Hamiltonian cell with the Hamiltonian density sandwiched
locate on the same cell; (ii) the hole cell locates on the right
hand side of the Hamiltonian cell; (iii) the  hole cell locates on
the left hand side of the Hamiltonian cell. In both cases (ii) and
(iii), there are $m$ cells between the hole cell and the Hamiltonian
cell, where $m \in (0,1,2,3,\cdots)$. As such, the four-index tensor
$A_{\ell rd}^{s}$ is updated as follows,
\begin{align}
A_{\ell rd}^{s}=A_{\ell rd}^{s} - \delta \; \frac{\partial
E}{\partial A_{\ell rd}^{s}},
\end{align}
where $\delta$ denotes the step size during updating. We stress
that, for a two-leg spin ladder, we should update four different
four-index tensors $A_{\ell rd}^{s}$, $B_{\ell rd}^{s}$, $C_{\ell
ru}^{s}$, and $D_{\ell ru}^{s}$ simultaneously.

The above updating procedure yields new tensors $A_{\ell rd}^{s}$,
$B_{\ell rd}^{s}$, $C_{\ell ru}^{s}$, and $D_{\ell ru}^{s}$ for a
two-leg spin ladder. Repeating this procedure until the ground-state
energy per unit cell converges, we anticipate that  the system's
ground-state wave function is generated in the TN representation.

For a three-leg spin ladder, one should introduce four different
four-index tensers $A_{\ell rd}^{s}$, $B_{\ell rd}^{s}$, $E_{\ell
ru}^{s}$, and $F_{\ell ru}^{s}$, and two different five-index
tensers $C_{\ell rud}^{s}$ and $D_{\ell rud}^{s}$. Similarly, more
tensors are needed for a multi-leg spin ladder. However, the
algorithm is applicable to a multi-leg spin ladder, as long as the
memory is sufficient to store the TN tensors.

{\it The models.} As an illustration, we test the algorithm on the
infinite-size two-leg and three-leg Heisenberg ladder systems with
staggering dimerization.

The two-leg and three-leg Heisenberg spin ladders are, respectively,
described by the Hamiltonian
\begin{equation}\label {fitlader1}
   H=\sum_{\alpha=1, 2}\sum_{i}J_{i, \alpha} S_{i, \alpha}\cdot S_{i
   +1, \alpha}
   +J_{\bot}\sum_{i} S_{i, 1}\cdot S_{i, 2},
 \end{equation}
and
\begin{equation}\label {fitlader2}
   H=\sum_{\alpha=1, 2, 3}\sum_{i}J_{i, \alpha} S_{i, \alpha}\cdot  S_{i+1, \alpha}
   +J_{\bot}\sum_{i}(  S_{i, 1}\cdot  S_{i, 2}+ S_{i, 2}\cdot {S}_{i,
   3})
 \end{equation}
where $S_{i, \alpha}$ denotes the spin-$1/2$ Pauli operator at site
$i$ on the $\alpha$-th leg, and $J_{\bot}$ is the exchange
interaction coupling along the rungs.

In order to test our algorithm, we first consider a two-leg ladder.
Choose the exchange interaction coupling constant $J_{\bot}\in 0.2,
0.4, 0.6, 0.8, 1.0$, and the coupling constant in each chain $J_{i,
\alpha}=J$ ($J=1$) if $i+\alpha$ is odd, $J_{i, \alpha}=J'$ if
$i+\alpha$ is even, with $J'\in 0, 0.2, 0.4, 0.6, 0.8, 1.0$. In
Table ~\ref{Tab1}, we list our simulation results for the
ground-state energy per site, for different values of $J_{\bot}$ and
$J'$, with the truncation dimension up to $6$, against the
extrapolated infinite-size ground-state energy per site from
finite-size spin-$1/2$ ladders in Ref.\cite{EDtwoleg}. The fact that
they matches very well demonstrates that our TN algorithm for spin
ladders is reliable.
\begin{center}
\begin{table}
\begin{tabular}{|c|c|l|c|}
 \hline $~~J_{\bot}~~~$ & $~~~J'~~~$ & $~~~~~~{\rm N.~Flocke}~~~~~$ & $~~~~~{\rm Our~results}~~~~~$
 \\ \cline{1-4}
      &0.0 &$-0.3769744936$ & $-0.376974$\\ \cline{2-4}
      &0.2 &$-0.37929324  $ & $-0.379293$\\ \cline{2-4}
  0.2 &0.4 &$-0.386139    $ & $-0.386139$\\ \cline{2-4}
      &0.6 &$-0.39850     $ & $-0.398509$\\ \cline{2-4}
      &0.8 &$-0.4181      $ & $-0.418215$\\ \cline{2-4}
      &1.0 &$-0.4516      $ & $-0.451554$\\ \cline{1-4}
      &0.0 &$-0.3833562502$ & $-0.383356$\\ \cline{2-4}
      &0.2 &$-0.3868021   $ & $-0.386801$\\ \cline{2-4}
  0.4 &0.4 &$-0.39515     $ & $-0.395154$\\ \cline{2-4}
      &0.6 &$-0.4096      $ & $-0.409689$\\ \cline{2-4}
      &0.8 &$-0.4334      $ & $-0.432975$\\ \cline{2-4}
      &1.0 &$-0.4712      $ & $-0.491242$\\ \cline{1-4}
      &0.0 &$-0.39504841  $ & $-0.395048$\\ \cline{2-4}
      &0.2 &$-0.40060     $ & $-0.400597$\\ \cline{2-4}
  0.6 &0.4 &$-0.4117      $ & $-0.411752$\\ \cline{2-4}
      &0.6 &$-0.4304      $ & $-0.430514$\\ \cline{2-4}
      &0.8 &$-0.4617      $ & $-0.461940$\\ \cline{2-4}
      &1.0 &$-0.4994      $ & $-0.499637$\\ \cline{1-4}
      &0.0 &$-0.41356     $ & $-0.413564$\\ \cline{2-4}
      &0.2 &$-0.4226      $ & $-0.422680$\\ \cline{2-4}
  0.8 &0.4 &$-0.4397      $ & $-0.439913$\\ \cline{2-4}
      &0.6 &$-0.4674      $ & $-0.467553$\\ \cline{2-4}
      &0.8 &$-0.4995      $ & $-0.499617$\\ \cline{2-4}
      &1.0 &$-0.5354      $ & $-0.535502$\\ \cline{1-4}
      &0.0 &$-0.4431413845$ & $-0.443063$\\ \cline{2-4}
      &0.2 &$-0.4629      $ & $-0.463080$\\ \cline{2-4}
  1.0 &0.4 &$-0.4870      $ & $-0.487120$\\ \cline{2-4}
      &0.6 &$-0.5143      $ & $-0.514348$\\ \cline{2-4}
      &0.8 &$-0.5446      $ & $-0.544634$\\ \cline{2-4}
      &1.0 &$-0.5780034099$ & $-0.578035$\\ \cline{2-4} \hline
\end{tabular}
\caption{The extrapolated infinite-size ground-state energy per site
from finite-size spin-$1/2$ ladders in Ref.~\cite{EDtwoleg} vs. our
ground-state energy per site for the infinite-size two-leg
Heisenberg ladder. }\label{Tab1}
\end{table}
\end{center}
Second, we focus on critical points of the ladders with the
staggered dimerization $J_{i, \alpha}=J[1+(-1)^{i+\alpha}\delta]$,
which are the exchange interaction couplings along the $\alpha$-th
leg for the two-leg ladder ($\alpha=1,2$) and the three-leg ladder
($\alpha=1,2,3$). In addition, we choose $\delta=0.5$, and the
coupling constant $J$ to be unity ($J=1$). To this end, we need to
compute the fidelity per lattice site.
\begin{figure}
\begin{center}
\includegraphics[height=42mm,width=0.40\textwidth]{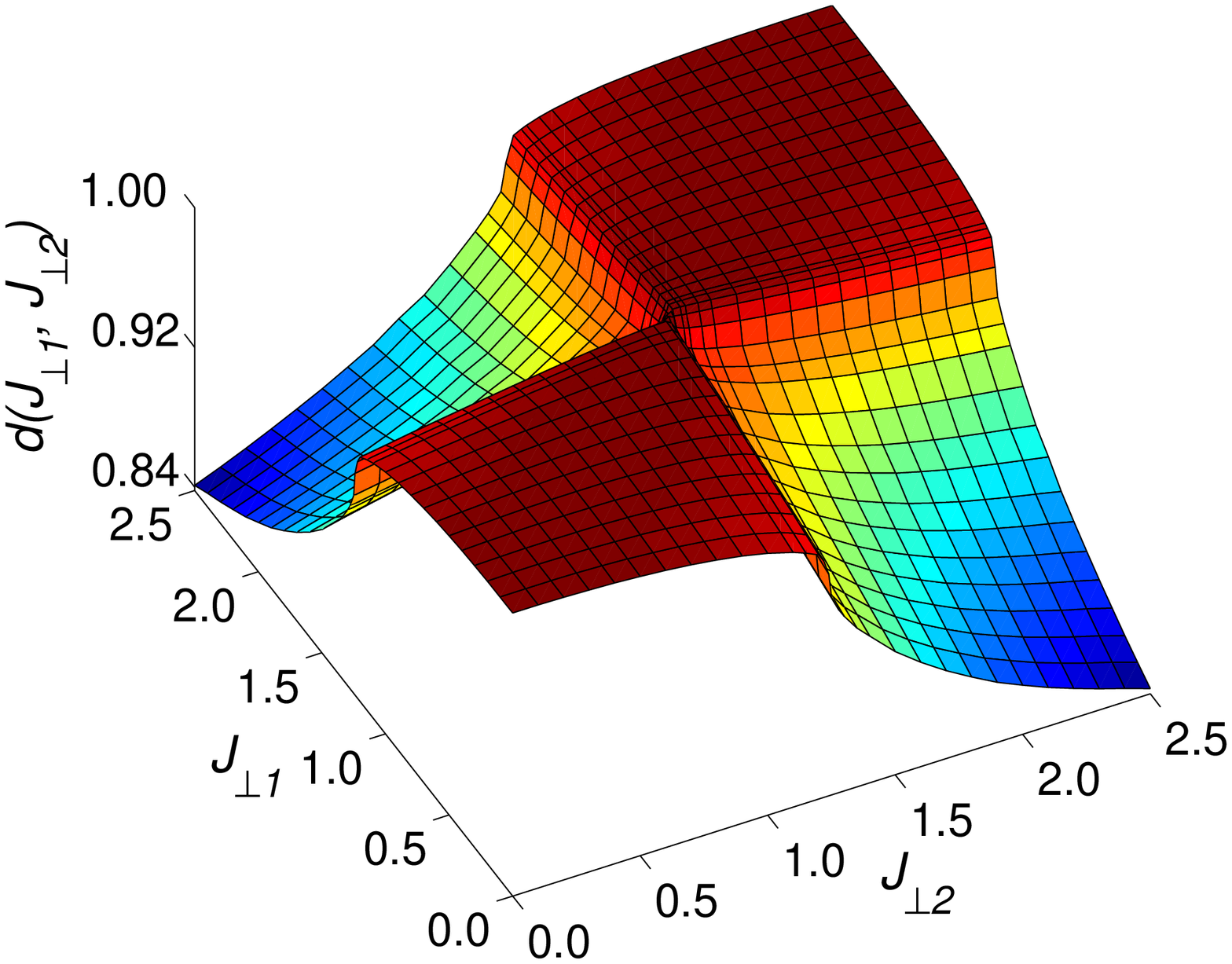}
\end{center}
\vspace*{0cm}
\begin{center}
\includegraphics[width=0.30\textwidth]{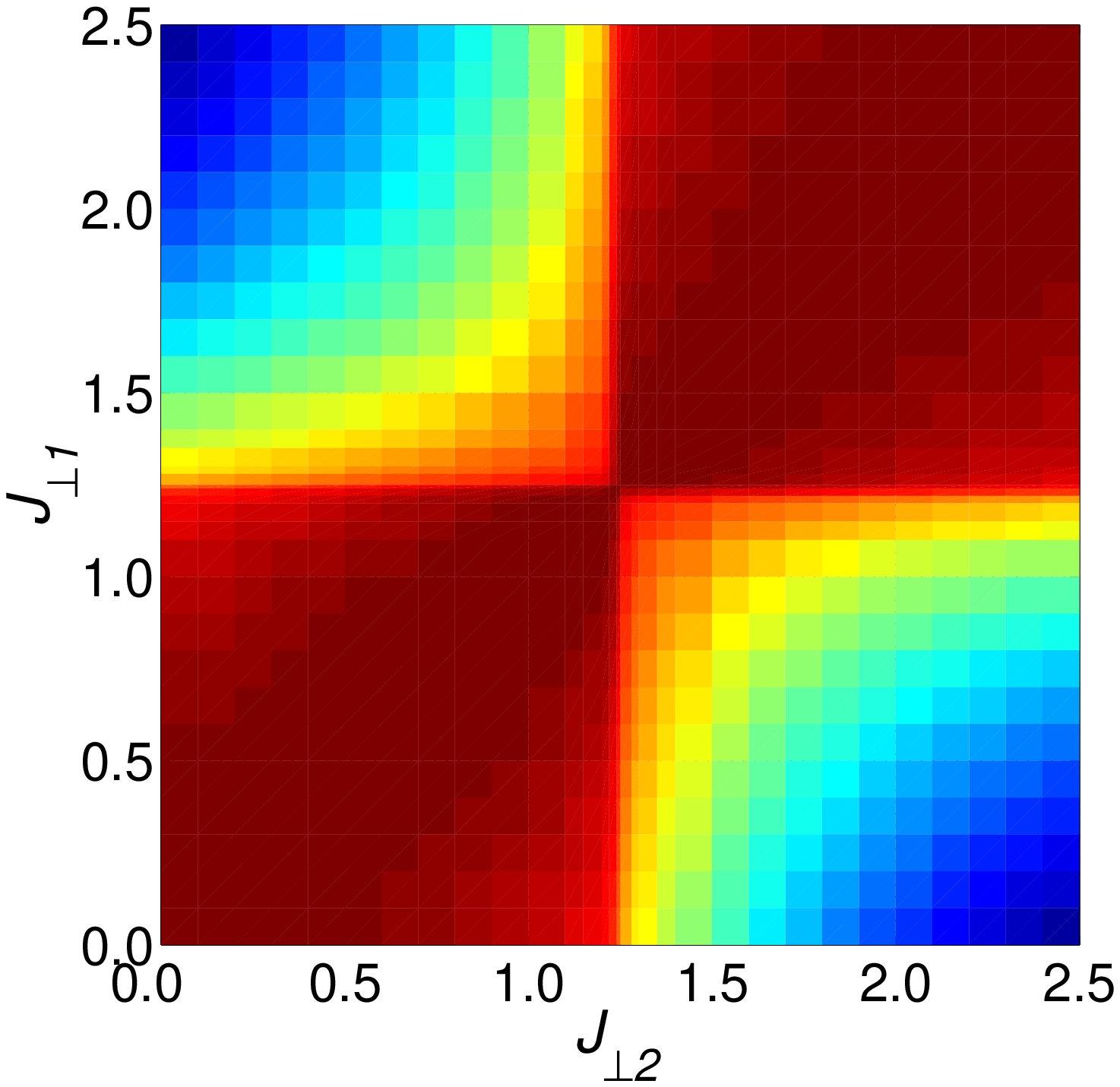}
\end{center}
\vspace*{0cm} \caption{(color online) The ground-state fidelity per
lattice site $d(J_{\bot1}, J_{\bot2})$, as a function of $J_{\bot1}$
and $J_{\bot2}$ for the two-leg Heisenberg ladder with staggering
dimerization. Upper panel: A two-dimensional fidelity surface
embedded in a three-dimensional Euclidean space. A continuous phase
transition point $J_{\bot c}\simeq1.24$ is identified as a pinch
point ($J_{\bot c}$,$J_{\bot c}$) on the fidelity surface, as argued
in Refs.~\cite{b1,b2,b3,b4}. Here, we have taken the truncation
dimension $\mathbb{D}=6$.  Lower panel: The contour plot of the
fidelity per lattice site $d(J_{\bot1}, J_{\bot2})$, on the
($J_{\bot1}, J_{\bot2}$)-plane, for  the two-leg Heisenberg ladder
with staggering dimerization.  } \label{leg2}
\end{figure}

\begin{figure}
\begin{center}
\includegraphics[height=42mm,width=0.40\textwidth]{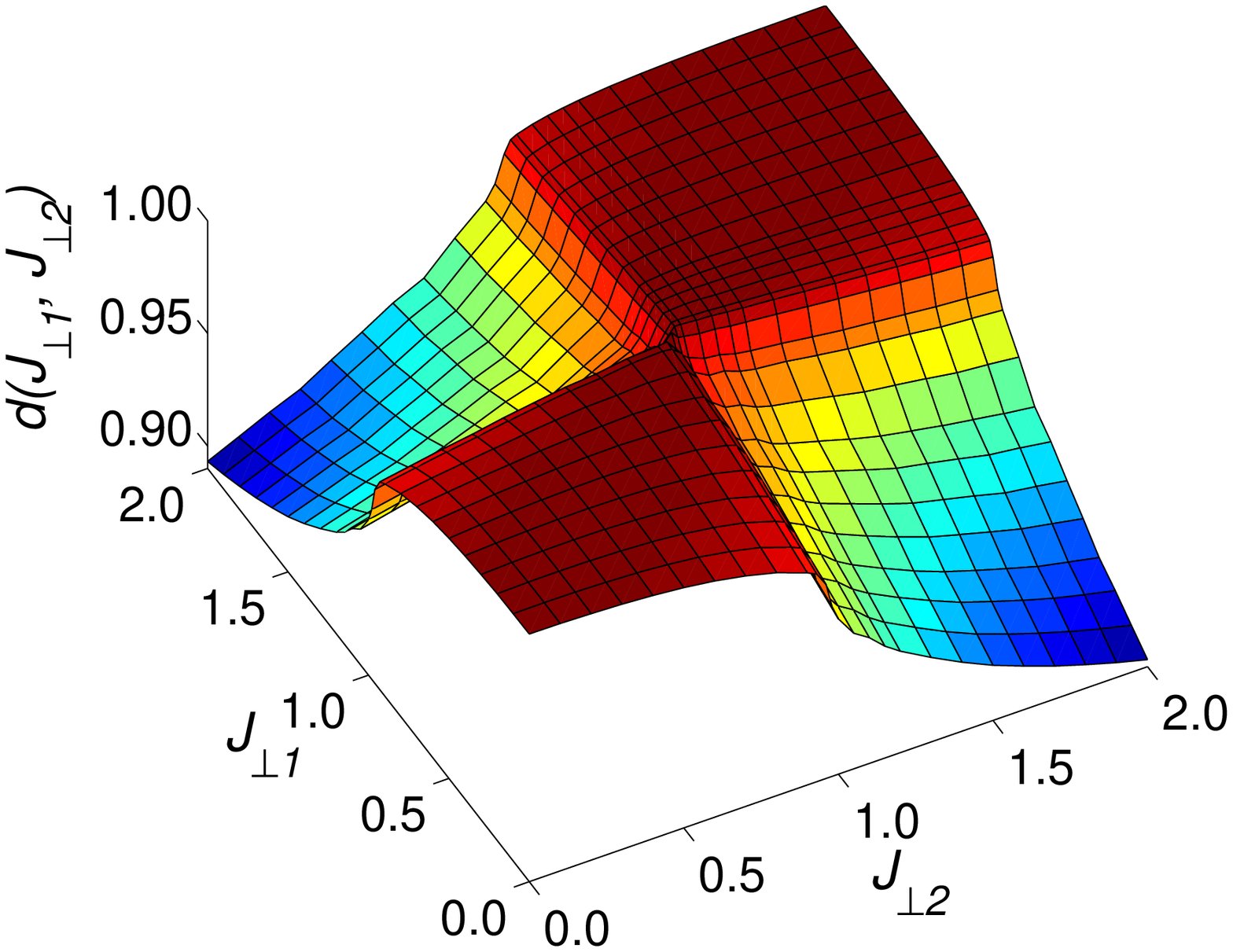}
\end{center}
\vspace*{0cm}
\begin{center}
\includegraphics[width=0.30\textwidth]{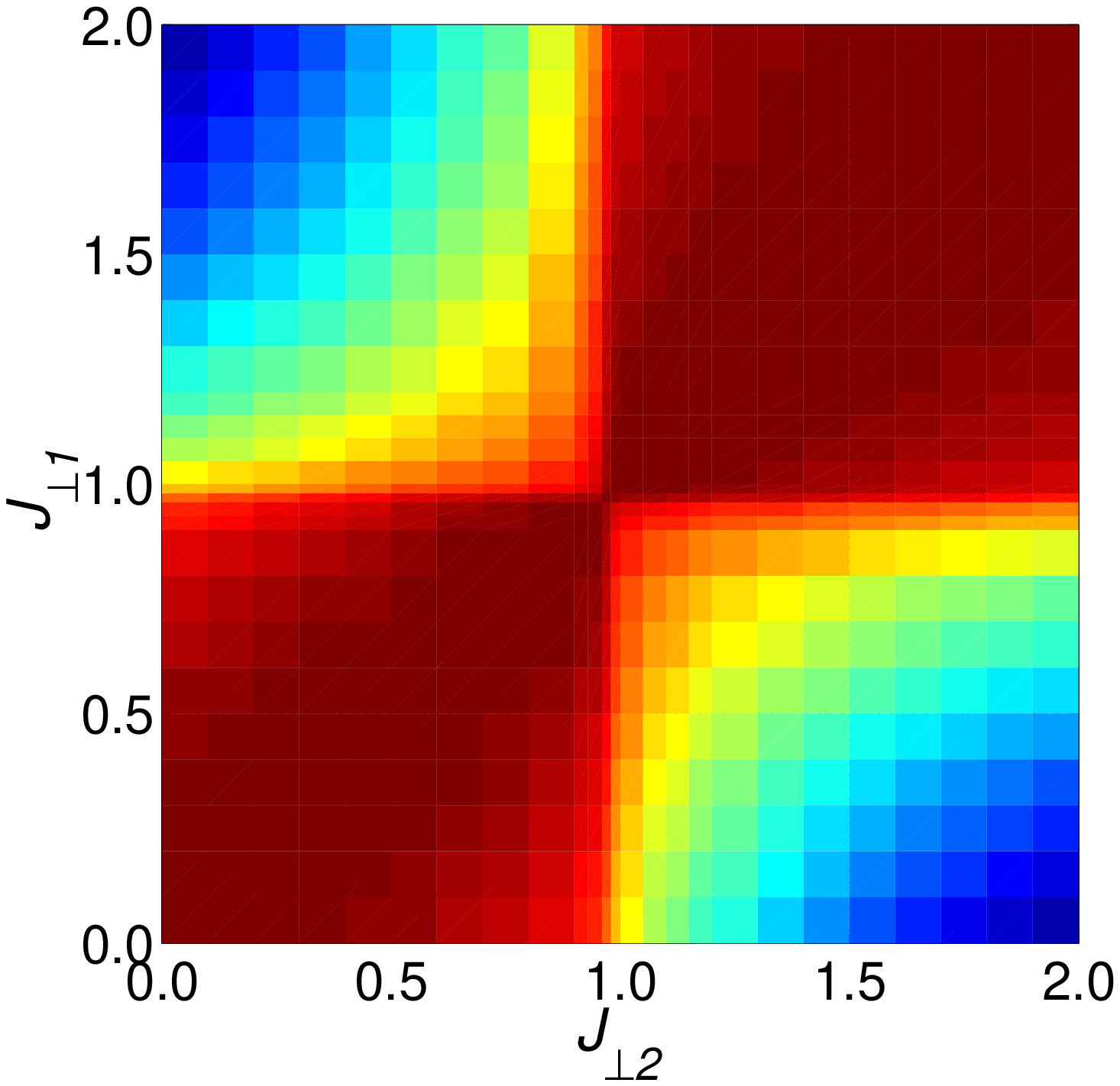}
\end{center}
\vspace*{0cm} \caption{(color online) The ground-state fidelity per
lattice site $d(J_{\bot1}, J_{\bot2})$, as a function of $J_{\bot1}$
and $J_{\bot2}$ for the three-leg Heisenberg ladder with staggering
dimerization. Upper panel: A two-dimensional fidelity surface
embedded in a three-dimensional Euclidean space. A continuous phase
transition point $J_{\bot c}\simeq 0.96$ is identified as a pinch
point ($J_{\bot c}$,$J_{\bot c}$) on the fidelity surface, as argued
in Refs.~\cite{b1,b2,b3,b4}. Here, we have taken the truncation
dimension $\mathbb{D}=6$.  Lower panel: The contour plot of the
fidelity per lattice site $d(J_{\bot1}, J_{\bot2})$, on the
($J_{\bot1}, J_{\bot2}$)-plane, for  the two-leg Heisenberg ladder
with staggering dimerization. } \label{leg3}
\end{figure}

{\it The ground-state fidelity per lattice site.} As argued in
Refs.~\cite{b1,b2,b3,b4}, the ground-state fidelity per lattice site
is a universal marker to detect a quantum phase transition: a phase
transition point is characterized by a pinch point on the fidelity
surface.

Consider the Heisenberg ladders with staggering dimerization. We
choose the exchange interaction coupling along the rungs $J_{\bot}$
as a control parameter. For two different ground states,
$|\psi(J_{\bot1})\rangle$ and $|\psi(J_{\bot2})\rangle$
corresponding to two different values $J_{\bot1}$ and $J_{\bot2}$ of
the control parameter $J_{\bot}$, the ground-state fidelity
$F(J_{\bot1},J_{\bot2})=|\langle\psi(J_{\bot2})|\psi(J_{\bot1})\rangle|$
asymptotically scales as $F(J_{\bot1},J_{\bot2})\sim
d(J_{\bot1},J_{\bot2})^N$, with $N$ the system size. Here,
$d(J_{\bot1},J_{\bot2})$ is the scaling parameter, introduced in
Refs.~\cite{b1,b2,b3} for one-dimensional quantum lattice systems
and in Ref.~\cite{b4} for two and higher-dimensional quantum lattice
systems; it characterizes how fast the fidelity between two ground
states goes to zero when the thermodynamic limit is approached.
Physically, the scaling parameter $d(J_{\bot1},J_{\bot2})$ is the
averaged fidelity per lattice site,
\begin{equation}
\ln d(J_{\bot1},J_{\bot2})\equiv \lim_{N\rightarrow \infty}\frac{\ln
F(J_{\bot1},J_{\bot2})}{N},
\end{equation}
which is seen to be well defined in the thermodynamic limit. It
satisfies the properties inherited from the fidelity
$F(J_{\bot1},J_{\bot2})$: (i) normalization $d(J_{\bot},J_{\bot})=
1$; (ii) symmetry $d(J_{\bot1},J_{\bot2})= d(J_{\bot2},J_{\bot1})$;
and (iii) range $0 \leq d(J_{\bot1},J_{\bot2}) \leq 1$.

We emphasize that the TN representation of the system's wave
functions generated from the algorithm makes it efficient to compute
the ground-state fidelity per lattice site for spin ladders.

In Fig.\ref{leg2}, we plot the ground-state fidelity per lattice
site for the two-leg Heisenberg spin-$1/2$ ladder with staggering
dimerization. A two-dimensional fidelity surface embedded in a
three-dimensional Euclidean space is shown in the upper panel, with
a pinch point at ($1.24$,$1.24$), implying that a continuous phase
transition occurs at $J_{\bot c}\simeq1.24$. In the lower panel, a
contour plot is shown for the fidelity per lattice site on the
($J_{\bot1}, J_{\bot2}$)-plane. We stress that no significant shifts
are observed for the pinch point, when  the truncation dimension is
increased up to $6$. Therefore, we conclude that a continuous phase
transition takes place at $J_{\bot c}\simeq1.24$, which is very
close to earlier results from the mean-field
theory~\cite{QCriticality}, exact diagonalization ~\cite{d2} and
DMRG~\cite{dimer}.

Similar to the two-leg Heisenberg spin-$1/2$ ladder, we  plot a two
dimensional fidelity surface embedded in a three-dimensional
Euclidean space, namely, the ground-state fidelity per lattice site
$d(J_{\bot1},J_{\bot2})$ as a function of $J_{\bot1}$ and
$J_{\bot2}$, for the three-leg Heisenberg spin-$1/2$ ladder with
staggering dimerization in Fig.~\ref{leg3}. It yields reliable
results, with only the truncation dimension up to $6$. A continuous
phase transition point $J_{\bot c}\simeq 0.96$ is identified as a
pinch point ($J_{\bot c}$, $J_{\bot c}$) on the fidelity surface,
consistent with the previous results from the DMRG
method~\cite{QCriticality,QCriticality3}.

{\it Conclusions.} We have developed an efficient TN algorithm to
compute ground-state wave functions for infinite-size quantum spin
ladders. Our investigation lends further support to the observation
that the ground-state fidelity per lattice site is able to
characterize critical phenomena in quantum many-body systems. It
also demonstrates that the developed TN algorithm for spin ladders
is efficient to compute the fidelity per lattice site.

{\it Acknowledgments.} We thank Bing-Quan Hu, Qian-Qian Shi, Bo Li,
Jin-Hua Liu, Hong-Lei Wang, and Jian-Hui Zhao for enlightening
discussions. This work is supported in part by the National Natural
Science Foundation of China (Grant No: 10874252). SHL, YHS and YWD
are supported by the Fundamental Research Funds for the Central
Universities (Project Nos: CDJXS11102213 and CDJXS11102214) and by
Chongqing University Postgraduates' Science and Innovation Fund
(Project No.: 200911C1A0060322).

\end{document}